\documentclass[english,12pt]{article}
\usepackage[LGR,T1]{fontenc}
\usepackage[utf8]{inputenc}
\usepackage{geometry}
\usepackage{array}
\usepackage{float}
\usepackage{booktabs}
\usepackage{textcomp}
\usepackage{multirow}
\usepackage{amsmath}
\usepackage{amsthm}
\usepackage{amssymb}
\usepackage{authblk}
\usepackage{sectsty}
\usepackage{hyperref}
\usepackage{graphicx}
\usepackage{blindtext}
\usepackage{caption}
\usepackage{subcaption}
\usepackage{lmodern}
\usepackage{multicol}

\usepackage{tikz}
\usetikzlibrary{arrows,automata}
\tikzset{
	semi/.style={
		semicircle,
		draw,
		minimum size=2em
	}
}

\usetikzlibrary{shapes, snakes, graphs, shapes.geometric, positioning}

\sectionfont{\fontsize{14}{13}\selectfont}
\subsectionfont{\fontsize{12}{12}\selectfont}

\allowdisplaybreaks

\makeatletter

\ProvideTextCommand{\~}{LGR}[1]{\char126#1}

\providecommand{\tabularnewline}{\\}

\theoremstyle{plain}

\usepackage{amsmath}

\makeatother

\usepackage{babel}
\providecommand{\theoremname}{Theorem}

\newcommand\independent{\protect\mathpalette{\protect\independenT}{\perp}}
\def\independenT#1#2{\mathrel{\rlap{$#1#2$}\mkern2mu{#1#2}}}

\usepackage[
backend=biber,
style=nature,
uniquename=init,
giveninits
]{biblatex}

\begin{document}

\title{Single-World Intervention Graphs for Defining, Identifying, and Communicating Estimands in Clinical Trials}


%


\author[1$\dag$]{Alex Ocampo}
\author[2]{Jemar R. Bather}

\affil[1]{Novartis Pharma AG, Basel, Switzerland}
\affil[2]{Department of Biostatistics, Harvard T.H. Chan School of Public Health, Boston, MA}

\affil[$\dag$]{Email: alex.ocampo@novartis.com}
\setcounter{Maxaffil}{0}
\renewcommand\Affilfont{\itshape\small}

\date{}
\maketitle
\vspace{-11mm}
\begin{abstract}
Confusion often arises when attempting to articulate target estimand(s) of a clinical trial in plain language. We aim to rectify this confusion by using a type of causal graph called the Single-World Intervention Graph (SWIG) to provide a visual representation of the estimand that can be effectively communicated to interdisciplinary stakeholders. These graphs not only display estimands, but also illustrate the assumptions under which a causal estimand is identifiable by presenting the graphical relationships between the treatment, intercurrent events, and clinical outcomes. To demonstrate its usefulness in pharmaceutical research, we present examples of SWIGs for various intercurrent event strategies specified in the ICH E9(R1) addendum, as well as an example from a real-world clinical trial for chronic pain. Latex code to generate all the SWIGs shown is this paper is made available. We advocate clinical trialists adopt the use of SWIGs in their estimand discussions during the planning stages of their studies. 
\end{abstract}



\section{Introduction}

The ICH E9(R1) addendum \cite{iche9} was a pivotal achievement in aligning all stakeholders (e.g., statisticians, clinicians, regulators) on target treatment effect(s) of randomized clinical trials (RCTs). This guidance came at an ideal time in statistical history as it aligned with work developed by pioneers in causal inference. Their  advances include methods to define, identify, and communicate causal treatment effects from both observational data and randomized trials \cite{robins2000marginal,robins1994estimation,pearl1995causal,hernan2004structural,robins1992identifiability,hernan2020causal,hernan2021methods}. Causal inferences has allowed us to define when we can step beyond the mantra, "correlation does not imply causation". This is pertinent to the pharmaceutical industry, where demonstrating causation of a new treatment is the primary objective.

While the tools of causal inference (e.g., potential outcomes, causal graphs, etc.) are not explicitly mentioned in the addendum, the language of causal inference appears throughout. For example, Section A.3 states that research questions such as ``how the outcome of treatment compares to what would have happened to the same subjects under alternative treatment'' are pivotal for drug development and licensing. Another example can be found in Section A.3.2: ``A scenario is envisaged in which the intercurrent event would not occur.'' These types of research questions and hypothetical scenarios can be defined using potential outcomes, prompting the application of causal inference methods. As a result, recent research has mathematically translated estimands from the ICH E9(R1) addendum into the causal inference framework \cite{lipkovich2020causal, mallinckrodt2019estimands}. This paper continues this trend by providing stakeholders with an accessible tutorial of how to characterize various estimands described in the addendum using causal graphs.

Causal graphs were popularized by Pearl \cite{pearl2009causality,greenland1999causal} in the form of Directed Acyclic Graphs (DAGs). Figure 1 shows a classic example of a DAG where a treatment $A$ has a causal effect on the intercurrent event $M$ and outcome $Y$. Here $M$ also has a causal effect on $Y$. Through a graphical representation, DAGs encode the research team's assumptions about which variables causally affect one another. DAGs make explicit the independencies (and conditional independencies) between variables, which can be read off of the graph using the $d$-separation criterion \cite{geiger1990d}. For these reasons, clinical trialists can use DAGs to effectively communicate the relationships among different variables of interest in the study. As a result, DAGs can serve as an invaluable tool to inform decision-making prior to performing statistical analyses. They also have the potential advantage of uncovering novel research questions within established trials, thus mitigating trial expenses while expanding therapeutic research.  

\begin{figure}[H]
    \begin{center}
    \begin{tikzpicture}[->,>=stealth']
    \tikzstyle{every state}=[draw=none]
    \node[shape=circle, draw] (A) at (0,0) {$A$};
    \node[shape=circle, draw] (M) at (2.5,0) {$M$};
    \node[shape=circle, draw] (Y) at (5,0) {$Y$};
    
      \path 
    	(A)  edge  [very thick, color=blue]                    (M)  
    	(A)  edge  [bend left,very thick, color=blue]  (Y)
    	(M)  edge  [very thick, color=blue]  (Y)						 
    	;
    \end{tikzpicture}
    \end{center}
    
        \caption{An example of a DAG with a treatment $A$, intercurrent event $M$, and outcome $Y$}
\end{figure}
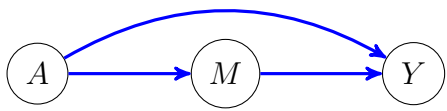

However, DAGs are not without limitation. Their primary shortcoming is that they do not visualize potential outcomes \cite{splawa1990application}, which are needed to define the estimands outlined in the ICH-E9(R1) addendum. In fact, as we will show in this paper, the DAG in Figure 1 could be associated with multiple estimands in a RCT, depending on the strategy used for handling intercurrent events. Luckily, Single-World Intervention Graphs (SWIGs) overcome this limitation by allowing DAGs to incorporate potential outcomes. As with DAGs, independencies among variables can be identified using $d$-separation. Using these properties, we demonstrate how to graphically represent common target estimands in clinical trials for various intercurrent event strategies using SWIGs. 

 This paper proceeds as follows. Section 2 provides a brief overview of SWIGs. This gives the reader the prerequisites to digest the example SWIGs in the subsequent sections. Section 3 provides examples of SWIGs for various intercurrent event strategies outlined in the ICH E9 addendum. Section 4 provides a case study of a SWIG in the context of a clinical trial for chronic pain. Lastly, Section 5 concludes with a discussion. 


\section{Overview of SWIGs}

In 2013, SWIGs were introduced by Richardson \& Robins to unite the potential outcome and graphical approaches to causal inference \cite{richardson2013single,richardson2013single2}. Their paper makes clear why SWIGs are necessary to do so, by considering the simplest DAG where treatment $A$ has a causal effect on clinical outcome $Y$,

\begin{figure}[H]

\begin{center}
\begin{tikzpicture}[->,>=stealth']
\tikzstyle{every state}=[draw=none]
\node [shape=circle,draw](A) at (0,0) {$A$};
\node[shape=circle,draw] (Y) at (3,0) {$Y$};

  \path 

	(A)  edge [very thick, color=blue]  (Y)
	;
\end{tikzpicture}
\end{center}

 \caption{Simplest DAG}
\end{figure}
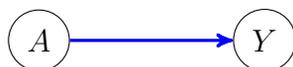

where interest lies in the estimand for the average treatment effect:
\[
\Delta = E[Y(a=1)] - E[Y(a=0)].
\]


Where $Y(a=1)$ is the potential outcome if experimental treatment were taken and $Y(a=0)$ is the potential outcome if reference treatment were taken. Note that all people in the population have both potential outcomes $Y(a=1)$ and $Y(a=0)$ despite that typically only one of them being observed in practice. For a review of potential outcomes framework, see \cite{rosenbaum1983central,rubin1978bayesian,holland1986statistics,robins1986new}. Clearly there is a disconnect between the DAG and the estimand, as the potential outcomes $Y(a=1)$ and $Y(a=0)$ do not appear on the DAG. This is where SWIGs provide added value. To see how, consider the following SWIGs associated with the DAG in Figure 2:

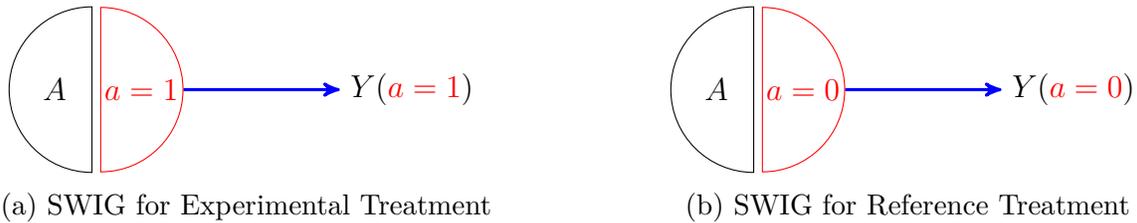
\begin{figure}[H]
\centering
\begin{subfigure}{.5\textwidth}
  \centering
  
  \begin{tikzpicture}[->,>=stealth']
\tikzstyle{every state}=[draw=none]
\node[shape=semicircle, draw, shape border rotate=90, inner sep=3.4mm] (A) at (0,0) {$A$};
\node[shape=semicircle, draw, shape border rotate=270, color=red, inner sep=0.5mm] (a) at (1.15,0) {$a=1$};
\node (Y) at (4.75,0) {$Y(\textcolor{red}{a=1})$};

  \path 
	(a)  edge  [very thick, color=blue]  (Y)
	;
\end{tikzpicture}

  \caption{SWIG for Experimental Treatment}
  \label{fig:sub1}
\end{subfigure}%
\begin{subfigure}{.5\textwidth}
  \centering

 \begin{tikzpicture}[->,>=stealth']
\tikzstyle{every state}=[draw=none]
\node[shape=semicircle, draw, shape border rotate=90, inner sep=3.4mm] (A) at (0,0) {$A$};
\node[shape=semicircle, draw, shape border rotate=270, color=red, inner sep=0.5mm] (a) at (1.15,0) {$a=0$};
\node (Y) at (4.75,0) {$Y(\textcolor{red}{a=0})$};

  \path 
	(a)  edge  [very thick, color=blue]  (Y)
	;
\end{tikzpicture}

  \caption{SWIG for Reference Treatment}
  \label{fig:sub2}
\end{subfigure}
\caption{Two possible SWIGs for DAG in Figure 2}
\label{fig:test}
\end{figure}

The SWIGs clearly display the potential outcomes $Y(a=1)$ and $Y(a=0)$. This is due to the node-splitting transformation that took variable $A$ on the DAG and split it into two components: $A$ and $a$. Splitting the node represents asking the same "what if" questions that the potential outcomes pose: "What would the outcome be in a world where everyone took experimental treatment ($a=1$) or reference treatment ($a=0$)?". Therefore, splitting the node on the DAG results in changing the downstream variable from the observed outcome $Y$ to its corresponding potential outcomes $Y(a=1)$ and $Y(a=0)$. Since in this scenario there are two possible potential outcomes, there are two possible SWIGs. In practice, we often collapse the possible SWIGs into one general SWIG for any arbitrary $a$, which is shown in Figure 4. 

\begin{figure}[H]
\centering

\begin{tikzpicture}[->,>=stealth']
\tikzstyle{every state}=[draw=none]
\node[shape=semicircle, draw, shape border rotate=90, inner sep=1.5mm] (A) at (0,0) {$A$};
\node[shape=semicircle, draw, shape border rotate=270, color=red, inner sep=2mm] (a) at (0.75,0) {$a$};
\node (Y) at (3,0) {$Y(\textcolor{red}{a})$};

  \path 
	(a)  edge  [very thick, color=blue]  (Y)
	;
\end{tikzpicture}
\caption{General graph representing the two SWIGs in Figure 3}
\end{figure}
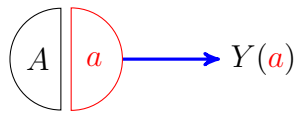

Most importantly, we can apply the same $d$-separation criterion used for DAGs to read independencies (and conditional independencies) between the variables from the SWIGs as well. In doing so we see that $A$ is independent of both $Y(a=1)$ and $Y(a=0)$ - i.e. $A \independent Y(a)$. This is because there is no backdoor path between $A$ and the potential outcomes on the SWIG. Also, note that the node-splitting blocks the forward path between $A$ and $Y(a)$. This is because we have now postulated a single-world where everyone took treatment $a$, so that the $A$ we observed in practice no longer effects $Y(a)$.

Given these independencies, we can identify the estimand of the average treatment effect $\Delta$ from the observed data as follows:

\begin{align*}
 \Delta &= E[Y(a=1)] - E[Y(a=0)]  \\    
                            & = E[Y(a=1)|A=1]-E[Y(a=0)|A=0] \hspace{7.5mm} (Y(a) \independent A ) \\
                            & = E[Y|A=1]-E[Y|A=0] \hspace{18mm} (\text{consistency})   
\end{align*}

The independency $Y(a) \independent A $, which was given from our assumed SWIG, allowed us to condition on $A$ in the expectation in line 2 of the equation. This is intuitive because if $A$ is independent of $Y(a)$ then we would expect the mean of $Y(a)$ to be the same in any subgroup of $A$ as in the whole population. In addition, we invoked the consistency assumption in the final line to change the potential outcome $Y(a)$ to the observed $Y$. The consistency assumption states that $Y=Y(a=1)A + Y(a=0)(1-A)$ \cite{cole2009consistency}. In layman's terms the consistency assumption states that the observed outcome under observed treatment is consistent with the underlying potential outcome for that same treatment. See, Table 1 for an example of consistency between potential outcomes and the observed data for five subjects - data that only God could see. Additionally, the usual causal assumptions of SUVTA \cite{rubin1980randomization} and positivity \cite{robins1986new} apply as well. 

\begin{table}[H]
\begin{centering}
\emph{}%
\begin{tabular}{ccccc}
\toprule 
id & $Y(a=0)$ & $Y(a=1)$ & $A$ & $Y$\tabularnewline
\midrule
\midrule 
1 & 60 & \textcolor{blue}{52} & 1 & \textcolor{blue}{52}\tabularnewline
\midrule 
2 & \textcolor{blue}{45} & 37 & 0 & \textcolor{blue}{45}\tabularnewline
\midrule 
3 & 46 & \textcolor{blue}{38} & 1 & \textcolor{blue}{38}\tabularnewline
\midrule 
4 & 75 & \textcolor{blue}{67} & 1 & \textcolor{blue}{67}\tabularnewline
\midrule 
5 & \textcolor{blue}{21} & 15 & 0 & \textcolor{blue}{21}\tabularnewline
\bottomrule
\end{tabular}
\par\end{centering}
\caption{God's Table}
\end{table}

Like DAGs, SWIGs encode one's assumptions about the process on which we are collecting data. It may be possible that two statisticians are involved in the same trial, but draw different SWIGs because they have different causal assumptions about the real world process. There may be differences in two statisticians choice on which nodes to split as well. However, the node-splitting decisions reflect interest in different estimands (or clinical questions) rather than causal assumptions about variables. This motivates the following section, which demonstrates SWIGs in the context of the ICH E9(R1) Addendum.


\section{Intercurrent Event Strategies from the ICH-E9(R1) Addendum in SWIGs}

We now explore visualizing various strategies for handling intercurrent events (IEs) using SWIGs. There are infinite ways to explore such estimands. Important work has started taking place in recent years to characterize specific clinical scenarios using SWIGs that incorporate time, multiple outcome measurements, and competing risks on the graph \cite{young2020causal,stensrud2021translating, parra2021hypothetical,breskin2018practical}. We have chosen to focus on simple cases in this section for pedagogical reasons in hopes that readers can afterwards extend the basic principles detailed in this paper to their own trial work. 

In all subsequent examples, $A$ is a binary treatment indicator that can be set to $a \in \{0,1\}$. We will consider a world where we intervene on $A$; consequently, this node will always be split in all subsequent SWIGs and all descendants of $a$ will take on their potential outcome representations. Also, because we are operating in the context of a randomized trial, $A$ will have no parents in all SWIGs presented; treatment is purely consequence of randomization and nothing else. The potential outcome $M(a)$ represents the value of the intercurrent event in a world where treatment $a$ is taken. We consider $M(a)=1$ if the intercurrent event occurs and $M(a)=0$ if not. Lastly, $Y(a)$ is the clinical outcome of interest in world where treatment $a$ is taken.  

\subsection{Treatment Policy}

\begin{figure}[H]
   
\begin{center}
\begin{tikzpicture}[->,>=stealth']
\tikzstyle{every state}=[draw=none]
\node[shape=semicircle, draw, shape border rotate=90, inner sep=1.5mm] (A) at (0,0) {$A$};
\node[shape=semicircle, draw, shape border rotate=270, color=red, inner sep=2mm] (a) at (0.75,0) {$a$};
\node (M) at (3,0) {$M(\textcolor{red}{a})$};
\node (Y) at (5.5,0) {$Y(\textcolor{red}{a})$};

  \path 
	(a)  edge  [very thick, color=blue]                    (M)  
	(a)  edge  [bend left,very thick, color=blue]  (Y)
	(M)  edge  [very thick, color=blue]  (Y)						 
	;
\end{tikzpicture}
\end{center}
    
    \caption{SWIG for an ITT Estimand}
    
\end{figure}
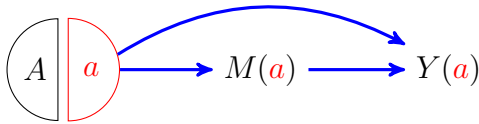

In the treatment policy, or intention to treat (ITT), strategy the intercurrent event is ignored. The corresponding SWIG (Figure 2) admits to this, as there is no node-splitting taking place on $M(a)$. The treatment effect (i.e. estimand) of interest measures a combination of the direct effect of the drug on the outcome and the indirect effect mediated through the intercurrent event. In the SWIG above, we can see that $Y(a)$ is independent of $A$, i.e. there are no paths between $A$ and $Y(a)$. Therefore, we can condition on $A$ in the expectations and then use the consistency assumption to identify the estimand from the observed data as follows:

\begin{align*}
 \Delta_{ITT} &= E[Y(1)] - E[Y(0)]  \\    
                            & = E[Y(1)|A=1]-E[Y(0)|A=0] \hspace{7.5mm} (Y(a) \independent A ) \\
                            & = E[Y|A=1]-E[Y|A=0] \hspace{18mm} (\text{consistency})   
\end{align*}

Identification is complete when the estimand of interest can be expressed as a function of the observed data alone \cite{pearl2022external,bareinboim2016causal} - i.e., no potential outcome variables remain in the expression. Note that for the ITT estimand above, the equivalence $E[Y(1)] - E[Y(0)] = E[Y(1,M(1))] - E[Y(0,M(0))]$ exists, which clearly demonstrates that the intercurrent event is ignored in this context. Due to this fact, it would actually be possible to collapse this SWIG into one identical to Figure 4 and retain the necessary independencies for identification; however, we recommend including even those IE that are ignored on the graph because this helps communicate that an ITT strategy has been adopted. 

\subsection{Hypothetical}

The hypothetical strategy envisages a counterfactual scenario - e.g., one in which the intercurrent event would not occur. A SWIG can make this explicit by splitting the node on the intercurrent event. Ergo, the "single world" that the SWIG depicts is one where, in addition to treatment being set to $a$, the intercurrent event is set to $m=0$. As a consequence of the node-splitting, the potential outcomes of interest are now $Y(a,m=0)$. This hypothetical estimand can be written mathematically as:

\begin{align*}
 \Delta_{hypo} &= E[Y(a=1,m=0)] - E[Y(a=0,m=0)]  \\    
\end{align*}

We present two examples of SWIGs for a hypothetical estimand. In this first example (Figure 6), there is an unobserved confounder $U$ between $M(a)$ and $Y(a,m)$. We indicate that this variable is unobserved by shading in the node.

\begin{figure}[H]
\begin{center}
\begin{tikzpicture}[->,>=stealth',scale=0.7]
\tikzstyle{every state}=[draw=none]

\node[shape=semicircle, draw, shape border rotate=90, inner sep=1.5mm] (A) at (0,0) {$A$};
\node[shape=semicircle, draw, shape border rotate=270, color=red, inner sep=2mm] (a) at (1,0) {$a$};

\node[shape=semicircle, draw, shape border rotate=90, inner sep=.1mm] (M) at (5,0) {$M(\textcolor{red}{a})$};
\node[shape=semicircle, draw, shape border rotate=270, color=red, inner sep=2.92mm] (m) at (6.5,0) {$m$};

\node (Y) at (11,0) {$Y(\textcolor{red}{a},\textcolor{red}{m})$};
\node[shape=circle,draw,fill=gray!40] (U) at (1,3) {$U$};

  \path 
	(a)  edge  [very thick, color=blue]                    (M)  
	(a)  edge  [bend right,very thick, color=blue,out=-45,in=-135]  (Y)
	(m)  edge  [very thick, color=blue]  (Y)
	(U) edge    [very thick, color=blue]         (M)
	(U) edge    [bend left, very thick, color=blue]     (Y)							 
	;
\end{tikzpicture}
\end{center}
    \caption{SWIG for a hypothetical estimand with an unobserved confounder}
    
\end{figure}
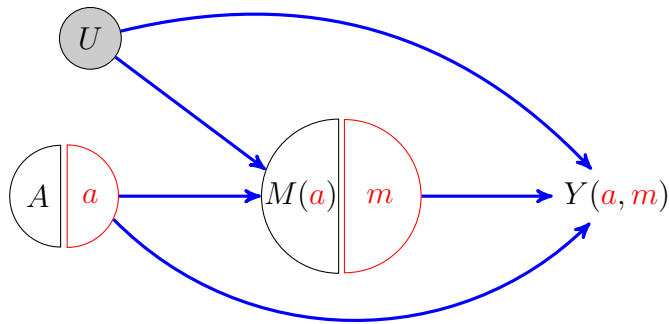

\begin{flushleft}
As we can see, $Y(a,m) \not\independent M(a) $ because of the backdoor path through $U$. In other words, when two variables share a common cause they are inherently related. Therefore, 
\end{flushleft}
\vspace{-5mm}
\begin{align*}
     E[Y(a=1,m=0)] - E[Y(a=0,m=0)] \neq E[Y|A=1,M=0] - E[Y|A=0,M=0]  
\end{align*}

Our attempts to identify the estimand fail because we cannot condition on the observed $M$ in the expectations. Consequently, we cannot express this causal contrast of potential outcomes as a function of the observed data. This example highlights the extra layer of information that the SWIG adds: the causal assumptions that determine whether identification of the estimand using the observed data is possible. SWIGs go one step beyond defining estimands in plain language or with tables.

In the second example, we assume to have observed a rich enough set of confounders between $M(a)$ and $Y(a,m)$ to obtain the required independence to identify the causal effect, i.e. the estimand, of interest. We denote conditioning on a variable, or blocking the causal pathway, by drawing a square box around this variable on the graph  (Figure 4).

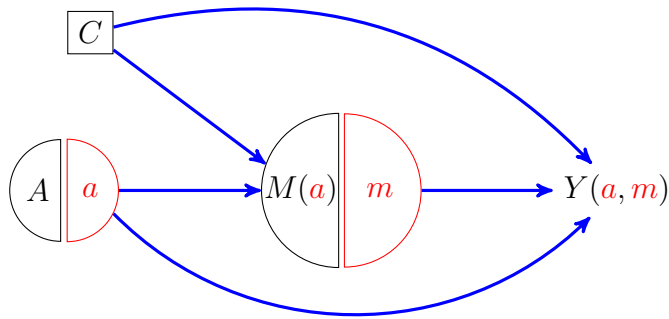
\begin{figure}[H]
\begin{center}
\begin{tikzpicture}[->,>=stealth',scale=0.7]
\tikzstyle{every state}=[draw=none]

\node[shape=semicircle, draw, shape border rotate=90, inner sep=1.5mm] (A) at (0,0) {$A$};
\node[shape=semicircle, draw, shape border rotate=270, color=red, inner sep=2mm] (a) at (1,0) {$a$};

\node[shape=semicircle, draw, shape border rotate=90, inner sep=.1mm] (M) at (5,0) {$M(\textcolor{red}{a})$};
\node[shape=semicircle, draw, shape border rotate=270, color=red, inner sep=2.92mm] (m) at (6.5,0) {$m$};

\node (Y) at (11,0) {$Y(\textcolor{red}{a},\textcolor{red}{m})$};
\node[shape=rectangle, draw] (C) at (1,3) {$C$};

  \path 
	(a)  edge  [very thick, color=blue]                    (M)  
	(a)  edge  [bend right,very thick, color=blue,out=-45,in=-135]  (Y)
	(m)  edge  [very thick, color=blue]  (Y)
	(C) edge    [very thick, color=blue]         (M)
	(C) edge    [bend left, very thick, color=blue]     (Y)							 
	;
\end{tikzpicture}
\end{center}
    \caption{SWIG for a hypothetical estimand with no unobserved confounding}
    
\end{figure}

The SWIG shows that we have observed and controlled for $C$ and can therefore induce the conditional independence $Y(a,m) \independent M(a) | C $. This allows us to reach identification by stratifying, or integrating, across levels of $C$:

%
 
 \vspace{-5mm}
 \begin{align*}
E[Y(a,m)] &=  \sum_c E[Y(a,m)|C=c]P(C=c) \\    
 & = \sum_c E[Y(a,m)|C=c,M(a)=m]P(C=c) \hspace{20mm} (Y(a,m) \independent M(a) | C  ) \\
 & = \sum_c E[Y(a,m)|C=c,M(a)=m,A=a]P(C=c) \hspace{7.5mm} (Y(a,m) \independent A ) \\
 & = \sum_c E[Y|C=c,M=m,A=a]P(C=c) \hspace{24mm} (\text{consistency}) \\
\end{align*}

Substituting this into $\Delta_{hypo}$:

\begin{align*}
    \Delta_{hypo} &= E[Y(a=1,m=0)] - E[Y(a=0,m=0)]  \\    
    &= \sum_c E[Y|C=c,M=0,A=1]P(C=c) -\sum_c E[Y|C=c,M=0,A=0]P(C=c) 
\end{align*}

Therefore, we've demonstrated that the hypothetical estimand can be identified from the observed data under the scenario defined by the SWIG in Figure 7. In this example, and in most hypothetical scenarios, we are making the assumption of no unobserved confounding between the IE and outcome. This is why, oftentimes, the sister assumption, missing at random (MAR)\cite{rubin1976inference}, is utilized for imputing these "missing" potential outcomes using an imputation model that leverages baseline covariates $C$. Of course, these are untestable assumptions, which highlight the importance of conducting a sensitivity analyses when considering a hypothetical estimand. Lastly, those readers familiar with the mediation literature in causal inference will note that many hypothetical estimands are examples of controlled direct effects \cite{vanderweele2015explanation}.

Of course, it must be plausible that one could in theory imagine such a world. It is possible to postulate a world in which everyone in your RCT population either receives treatment or does not receive treatment. However, the ability to conceive of a world in which the IE never happens will also depend on the specific IE at hand. Depending on the clinical context, it may be plausible to consider a world in which rescue medication is not taken or treatment discontinuation due to administrative dropout did not occur. In the event of active drug side effects, this may be less plausible. In the case of death, this is highly implausible.

\subsection{Composite}

Composite estimands incorporate the IE into the variable definition. In the simplest case of a binary outcome $Y(a)\in \{0,1\}$ the composite variable is defined as:

\[
U(a)=\begin{cases}
Y(a) & \text{if    } M(a)=0\\
0 & \text{if    } M(a)=1\\
\end{cases}
\vspace{3mm}
\]

Which corresponds to treating occurrence of an IE as a failure. Moreover, there exist strategies for creating composite estimands for continuous endpoints \cite{permutt2017trimmed, ocampo2019identifying}. The composite strategy is often employed in the case where the IE $M(a)$ under consideration is death, and hence the potential outcome $Y(a)$ doesn't exist when the IE occurs (i.e. $M(a)=1$). We characterize such an example in Figure 8, which assumes no causal effect of the IE on outcome and omit the arrow from $M(a)$ to $Y(a)$. An example of a SWIG for such a composite estimand could then be:

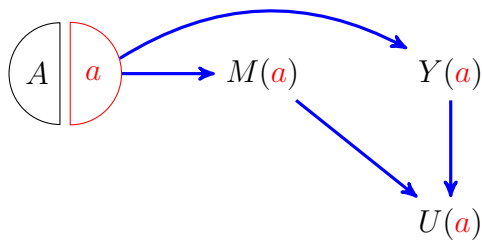
\begin{figure}[H]
    
\begin{center}
\begin{tikzpicture}[->,>=stealth']
\tikzstyle{every state}=[draw=none]
\node[shape=semicircle, draw, shape border rotate=90, inner sep=1.5mm] (A) at (0,0) {$A$};
\node[shape=semicircle, draw, shape border rotate=270, color=red, inner sep=2mm] (a) at (0.75,0) {$a$};
\node (M) at (3,0) {$M(\textcolor{red}{a})$};
\node (Y) at (5.5,0) {$Y(\textcolor{red}{a})$};
\node (U) at (5.5,-2) {$U(\textcolor{red}{a})$};

  \path 
	(a)  edge  [very thick, color=blue]                    (M)  
	(a)  edge  [bend left,very thick, color=blue]  (Y)	
        (M)  edge  [very thick, color=blue]  (U)						 
        (Y)  edge  [very thick, color=blue]  (U)
	;
\end{tikzpicture}
\end{center}
\caption{SWIG for a composite estimand}
    
\end{figure}

It is straightforward to see on the SWIG that $U(a)$ is a combination of the outcome $M(a)$ and $Y(a)$. It is also clear from the SWIG that $A \independent U(a)$, allowing us to identify the composite estimand:

\begin{align*}
 \Delta_{C} &= E[U(1)] - E[U(0)]  \\    
                            & = E[U(1)|A=1]-E[U(0)|A=0] \hspace{7.5mm} (U(a) \independent A ) \\
                            & = E[U|A=1]-E[U|A=0] \hspace{15mm} (\text{by consistency})   
\end{align*}

 
In the above example, we do not assume that the IE has a causal effect on the outcome - i.e. no arrow from $M(a)$ to $Y(a)$. This would be the case if the IE were death for example. There are of course situations where interest lies in a composite estimand, but there is an arrow from $M(a)$ to $Y(a)$. For example, this would unarguably be the case where $M$ is treatment discontinuation. Under such scenarios, the composite estimand remains identifiable. In fact, much like the ITT estimand it would remain identifiable even if there are unobserved confounders between $M(a)$ and $Y(a)$ because no matter what, the required independence $U(a) \independent A$ remains.

\subsection{Principal Stratum}

In the Principal Stratum (PS) strategy, interest lies in a sub-population of patients defined by whether or not an IE would occur under a particular treatment. We consider one such example: the population of patients who would not have the IE under active treatment, i.e. $M(a=1)=0$. This causal contrast, or estimand, of interest is therefore:
\vspace{-0mm}
\begin{align*}
\Delta_{PS} =E[Y(a=1)|M(a=1)=0] - E[Y(a=0)|M(a=1)=0]    
\end{align*}

Consider first the following SWIG, that considers a world in which every patient took the treatment and conditions on the strata of interest.

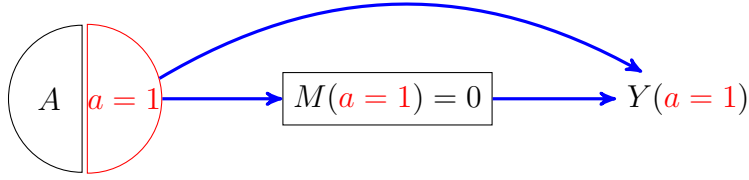
\begin{figure}[H]
\begin{center}
\begin{tikzpicture}[->,>=stealth']
\tikzstyle{every state}=[draw=none]
\node[shape=semicircle, draw, shape border rotate=90, inner sep=2.85mm] (A) at (0,0) {$A$};
\node[shape=semicircle, draw, shape border rotate=270, color=red, inner sep=.01mm] (a) at (1,0) {$a=1$};
\node[shape=rectangle, draw] (M) at (4.5,0) {$M(\textcolor{red}{a=1})=0$};
\node (Y) at (8.5,0) {$Y(\textcolor{red}{a=1})$};

  \path 
	(a)  edge  [very thick, color=blue]                    (M)  
	(a)  edge  [bend left,very thick, color=blue]  (Y)
	(M)  edge  [very thick, color=blue]  (Y)						 
	;
\end{tikzpicture}
\end{center}
\caption{SWIG where active treatment is given for those who would not have the IE under treatment (i.e. conditioning on $M(a=1)=0$)}
\end{figure}

The SWIG reveals two important facts. Firstly, as is true in all randomized trials $A \independent Y(a)$. Secondly, and most importantly, this makes clear that we are interested in the direct effect of treatment on outcome, as sub-setting to the principal strata $M(a=1)=0$ blocks the effect of treatment mediated through the IE. Using this SWIG, we can identify the expected value of $Y(a=1)$ in our principal strata, i.e. the first term in $\Delta_{PS}$:

\begin{align*}
 E[Y(a=1)|M(a=1)=0]  &=  E[Y(a=1)|M(a=1)=0,A=1] \hspace{2.5mm} (Y(a) \independent A ) \\
                 & = E[Y|M=0,A=1] \hspace{5mm} (\text{by consistency})   
\end{align*}

Which is straightforward as before. The difficulty in PS estimands comes into play when trying to identify the second term in $\Delta_{PS}$ from the observed data, i.e. $E[Y(a=0)|M(a=1)=0]$. If we consider the SWIG for the untreated world

\begin{figure}[H]
\begin{center}
\begin{tikzpicture}[->,>=stealth']
\tikzstyle{every state}=[draw=none]
\node[shape=semicircle, draw, shape border rotate=90, inner sep=2.85mm] (A) at (0,0) {$A$};
\node[shape=semicircle, draw, shape border rotate=270, color=red, inner sep=.01mm] (a) at (1,0) {$a=0$};
\node (M) at (4.5,0) {$M(\textcolor{red}{a=0})$};
\node (Y) at (8.5,0) {$Y(\textcolor{red}{a=0})$};

  \path 
	(a)  edge  [very thick, color=blue]                    (M)  
	(a)  edge  [bend left,very thick, color=blue]  (Y)
	(M)  edge  [very thick, color=blue]  (Y)						 
	;
\end{tikzpicture}
\end{center}
\caption{SWIG where reference treatment is given for those who would not have the IE under treatment $M(a=1)=0$}
\end{figure}
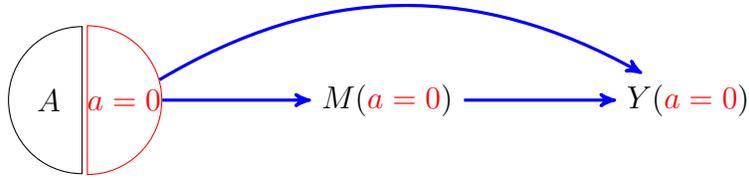

it becomes evident that identifying the second term requires a cross-world assumption. We want to know the expected value of $Y(a=0)$ in the strata of $M(a=1)=0$, but these variables do not appear on the same SWIG, or in the same world. This is a fundamental challenge of PS estimands and SWIGs help to make this challenge explicit when communicating the estimand. Of course, this should not downplay the clinical importance of trying to understand PS estimands and has not hindered statisticians attempts to estimate them \cite{bornkamp2021principal}. The SWIG helps makes explicit the assumptions needed to do so, and then allows us to begin discussing how plausible PS assumptions may be. 

Given the independencies presented on the two SWIGs, consider trying to identify the second term of this principal stratum estimand from the observed data:

\begin{align*}
 E[Y(a=0)|M(a=1)=0] &= E[Y(a=0)|M(a=1)=0,A=0] \hspace{2.5mm} (Y(a) \independent A ) \\
                &= E[Y|M(a=1)=0,A=0] \hspace{5mm} (\text{by consistency}) 
\end{align*}

The challenge arises in identifying the stratum of patients in the reference arm of the trial who would not have had the IE had they taken experimental treatment, something we never see in practice. The SWIG can help us to communicate this. 


\section{Example from a Clinical Trial for Chronic Pain}

We now outline an example of how one can use a SWIG to characterize an estimand in a Phase II clinical trial for chronic pain. Callegari et al. \cite{callegari2020estimands} provide a detailed exposition of the various attributes necessary to define a slew of clinically relevant target estimands in this context. Herein, we focus on the primary estimand outlined in their paper. We demonstrate how a SWIG can be used to define this estimand. This SWIG then allows the independencies between variables to be read off this graph to demonstrate identification of the estimand. This case study illustrates how a SWIG can aid in communicating a target estimand from a clinical trial.

To construct the graph, consider the following notation for the relevant variables used to characterize the estimand so that we can concisely place them on the graph. As above, we denote $A$ as the random variable representing treatment assignment. In the context of a chronic pain outlined in Callegari et al, this variable can take on two values: $A=1$ for active treatment and $A=0$ for placebo. However, because the SWIGs are equivalent under either treatment, we only need to draw one SWIG wherein we intervene on $A$ and set the value to arbitrary $a\in \{0,1\}$. We denote the relevant intercurrent events as:

\begin{itemize}
    \item $M_1$ = Intake of short acting pain relief medication
    \item $M_2$ = Treatment discontinuation due to Adverse Event, Loss of Efficacy, or intake of prohibited medications
    \item $M_3$ = Change of dose of allowed concomitant medication for pain
    \item $M_4$ =  Treatment discontinuation due to Administrative or Other reasons
\end{itemize}

Their primary estimand definition outlines the IE strategy used for each of the intercurrent events. That is, $M_1$ and $M_2$ a treatment policy strategy is adapted. That is, the data post IE is deemed relevant for these IE and collected and used in the analysis. For IEs $M_3$ and $M_4$, a hypothetical strategy is most relevant. Data post IE, even if available, are discarded and treated as missing. For these IEs, what would have happened had the IEs not occured (i.e. $M_3=M_4=0$) is most clinically relevant. The potential outcome of interest is $Y(a,m_1=0,m_4=0)$, representing weekly mean pain score change from baseline using the 11-point numerical rating scale that would be observed had treatment $a$ been taken, and had intercurrent events $M_1$ and $M_4$ not occurred. Thus, the estimand of interest is:

\[
\Delta_{RCT} = E[Y(a=1,m_1 = 0, m_4=0)] - E[Y(a=0,m_1 = 0, m_4=0)] 
\]

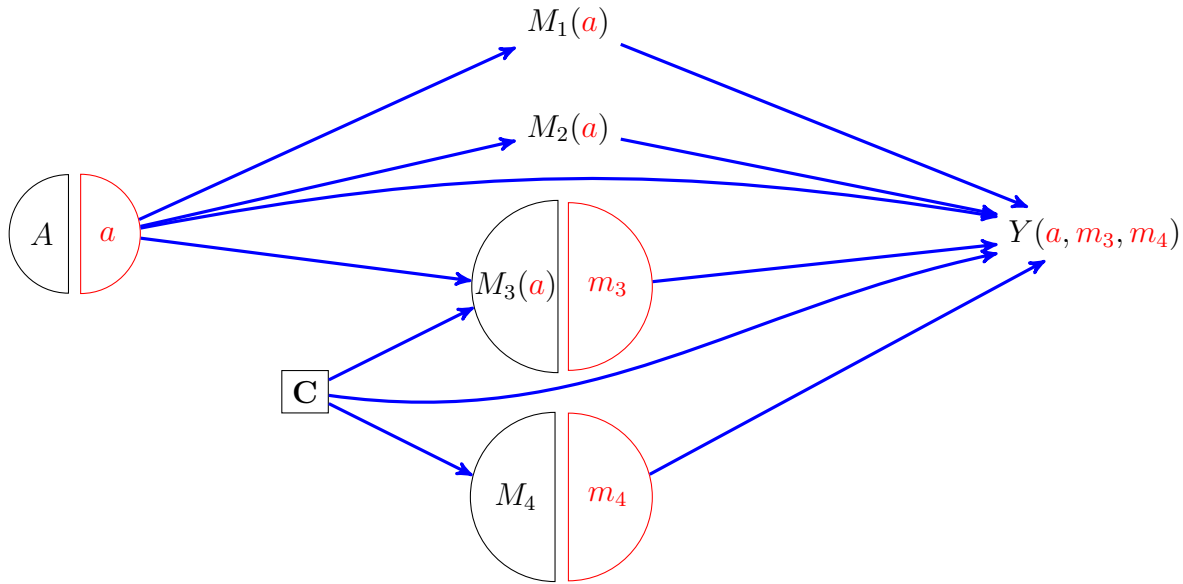
\begin{figure}[H]
\begin{center}
\begin{tikzpicture}[->,>=stealth',scale=0.7]
\tikzstyle{every state}=[draw=none]


\node[shape=semicircle, draw, shape border rotate=90, inner sep=2mm] (A) at (0,0) {$A$};
\node[shape=semicircle, draw, shape border rotate=270, color=red, inner sep=2.5mm] (a) at (1.25,0) {$a$};

\node (M1) at (10,4) {$M_1(\textcolor{red}{a})$};
\node (M2) at (10,2) {$M_2(\textcolor{red}{a})$};

\node[shape=semicircle, draw, shape border rotate=90, inner sep=.1mm] (M3) at (9,-1) {$M_3(\textcolor{red}{a})$};
\node[shape=semicircle, draw, shape border rotate=270, color=red, inner sep=2.6mm] (m3) at (10.75,-1) {$m_3$};

\node[shape=semicircle, draw, shape border rotate=90, inner sep=2.4mm] (M4) at (9,-5) {$M_4$};
\node[shape=semicircle, draw, shape border rotate=270, color=red, inner sep=2.6mm] (m4) at (10.75,-5) {$m_4$};

\node (Y) at (20,0) {$Y(\textcolor{red}{a},\textcolor{red}{m_3},\textcolor{red}{m_4})$};

\node[shape=rectangle, draw] (C) at (5,-3) {$\textbf{C}$};

  \path 
	(a)  edge  [very thick, color=blue]                    (M1)  
	(a)  edge  [very thick, color=blue]                    (M2)  
	(a)  edge  [bend left,very thick, color=blue,out=10,in=170]  (Y)
	(a)  edge  [very thick, color=blue]                    (M3)

	(M1)  edge  [very thick, color=blue]                    (Y)  
	(M2)  edge  [very thick, color=blue]                    (Y)  
	(m3)  edge  [very thick, color=blue]                    (Y)
	(m4)  edge  [very thick, color=blue]                    (Y)

	(C) edge    [very thick, color=blue]         (M3)
	(C) edge    [very thick, color=blue]         (M4)
	(C) edge    [bend right, very thick, color=blue,out=-20,in=180]     (Y)							 
	;
\end{tikzpicture}
\end{center}
    \caption{SWIG for a Clinical Trial in Chronic Pain}
    
\end{figure}

It is clear from the SWIG that $A$, $M_1$, and $M_4$ are d-separated from the potential outcome $Y(a,m_1,m_4)$ given that we can condition on $C$. These are the necessary independencies for identification. The mathematics for identifying this estimand are shown in supplementary information 1.





\section{Discussion}

It is easy to get lost in the jargon of estimands; SWIGs provide an escape. These causal graphs allow a quick view of the clinically relevant treatment effect. Such visuals may appeal to many, considering that roughly 65\% of the population are visual learners \cite{bradford2004reaching}. Time is of the essence to get effective treatments to those in need; tools like SWIGs help hasten estimand discussions. Additionally, SWIGs can limit mistakes and misalignment between stakeholders as they are succinct and clear. 

SWIGs encode clinical assumptions about how treatment, IE, confounders, and outcomes all casually interact with one another. As seen from the examples in this paper, these assumptions govern whether or not an estimand is identifiable, and thus estimable from the observed data. IE often embed observational studies within the RCT, and SWIGs help statisticians understand how to account for the biases. We recommend clinical trialists adopt SWIGs in preliminary planning discussions when defining estimands. The visuals are succinct and the rules of identification are simple to explain to interdisciplinary teams.

Furthermore, defining estimands explicitly using the mathematical notation of potential outcomes removes ambiguity. For instance, in the case of the clinical trial example in section 4, the proposed analysis leads to confusion about what potential outcome is being targeted. The authors consider treatment discontinuation due to adverse events, loss of efficacy, and intake of prohibited medications as unfavorable events in their estimand definition. In these scenarios, they propose using retrieved drop out (RDO) data in the analysis, which is aligned with a treatment policy strategy. That would imply interest in potential outcome $Y(a=1,m_1=0,m_4=0)$, which is to ignore $M_3$ as we did in section 4. However, should RDO data be unavailable, the authors suggest imputing outcomes based on the placebo data, which rather implies interest in $Y(a=0,m_1=0,m_4=0)$ among the principal stratum $M_3(a=1)=1$. Perhaps it is believed that these potential outcome distributions are equivalent. Nevertheless, the clinically relevant estimand needs to be made clear. Furthermore, their primary estimand definition written in section 3.1 does not align with their characterization of the estimand in Fig 1. This is because the latter proposes a different IE strategy for $M_2$ in the placebo arm, which was not mentioned in section 3.1. The SWIG removes these ambiguities from the estimand definition by clearly showing the potential outcome of interest on the graph.  

Our work is limited in a number of ways. Firstly, we ignore the "while on treatment" IE strategy in our presentation of SWIGs because of our own bias that this type of estimand is difficult to interpret causally. Additionally, previous work detailing causal diagrams in the estimand framework has added another mediating variable between the IE and outcome representing the intervention take post IE \cite{lipkovich2020causal}, but we find this redundant as their relationship is often deterministic and can be considered jointly in one variable. This may be a simplification on our part. Should the distinction prove valuable, this or any other variables should be added to the graph. Our main goal herein is to show the utility of SWIGs in characterizing estimands. In practice, many nuances specific to the trial will reveal themselves, making the SWIGs more complicated than those presented in this paper. Lastly, we used differences in expectations throughout to define our estimands for simplicity, but causal estimands can be contrasts (e.g. ratios, etc.) of any marginal statistical functional (e.g. hazards, odds, etc.). 


It is important for clinical trialists--especially statisticians--to learn to speak the language of causal inference in order to adapt the vast methodological progress made in this area over the last few decades. We believe incorporating SWIGs into estimand discussions is one such step in the right direction. To enable our readers to start using SWIGs, all LaTeX code to generate the graphs in this paper in included in supporting information 2.

        


\newpage
\printbibliography

\newpage

\begin{flushleft}
\textbf{Appendix: Identification of Estimand in Clinical Trial Example}
\end{flushleft}

As stated in section 4, the causal estimand of interest is:

\[
\Delta_{RCT} = E[Y(a=1,m_3 = 0, m_4=0)] - E[Y(a=0,m_3 = 0, m_4=0)] 
\]

Using the independencies on the SWIG in Figure X we can identify the estimand. Since the procedure for identifying the two expectations of potential outcomes in the contrast above are identical, we show identification of arbitrary $E[Y(a,m_3 = 0, m_4=0)]$:

\begin{align*}
E[Y(a,m_3 = 0, m_4=0)] &= E[Y(a,m_3 = 0, m_4=0)|A=a] \\
&= \sum_c E[Y(a,m_3 = 0, m_4=0)|A=a,C=c]Pr(C=c) \\
&= \sum_c E[Y(a,m_3 = 0, m_4=0)|A=a,C=c,M_3=0,M_4=0]Pr(C=c) \hspace{2.5mm} (1) \\
&= \sum_c E[Y|A=a,C=c,M_3=0,M_4=0]Pr(C=c) \hspace{2.5mm} (2) 
\end{align*}

Which leads to identification of the estimand. We use the fact that $\{M_3,M_4\} \independent Y(a,m_3 = 0, m_4=0) | C$ in line (1) to invoke the consistency assumption in line (2) and identify the causal effect from the observed data.

\end{document}